\documentclass[notitlepage,onecolumn,showpacs,11pt]{revtex4-1}
\usepackage[utf8]{inputenc}
\usepackage{amsmath}
\usepackage{amsfonts}
\usepackage{amssymb}
\usepackage{times,fullpage}
\usepackage{comment}
\usepackage{array,bm}
\usepackage{graphicx,tikz}

\begin{document}

\title{Two-dimensional ASEP model to study density profiles in CVD growth}
\author{Gagan Kumar} 
 \author{Annwesha Adhikari}
 \author{Anupam Roy}
 \email{royanupam@bitmesra.ac.in}
\author{Sourabh Lahiri}
\email{sourabhlahiri@bitmesra.ac.in}
\affiliation{Department of Physics, BIT Mesra, Ranchi, Jharkhand 835215, India}

\begin{abstract}
The growth of two-dimensional (2D) transition metal dichalcogenides using chemical vapor deposition has been an area of intense study, primarily due to the scalability requirements for potential device applications. One of the major challenges of such growths is the large-scale thickness variation of the grown film. To investigate the role of different growth parameters computationally, we use a 2D asymmetric simple-exclusion process (ASEP) model with open boundaries as an approximation to the dynamics of deposition on the coarse-grained lattice. 
    The variations in concentration of particles (growth profiles)  at the lattice sites  in the grown film  are studied as functions of parameters like injection and ejection rate of particles from the lattice, time of observation, and the right bias (difference between the hopping probabilities towards right and towards left) imposed by the carrier gas. In addition, the deposition rates at a given coarse-grained site is assumed to depend on the occupancy of that site. The effect of the maximum deposition rate, i.e., the deposition rate at a completely unoccupied site on the substrate, has been explored. 
    The growth profiles stretch horizontally when either the evolution time or the right bias is increased. An increased deposition rate leads to a step-like profile, with the higher density region close to the left edge. In 3D, the growth profiles become more uniform with the increase in the height of the precursor with respect to the substrate surface. These results qualitatively agree with the experimental observations. 
    
\end{abstract}

\maketitle

\section{\label{sec:Intro}Introduction}

Two-dimensional (2D) transition metal dichalcogenides (TMDs) have been focus of research owing to their many unique physical, electronic, and optoelectronic properties \cite{zeng2018exploring,xu2014spin,xiao2012coupled} that can lead to wide range of applications in electronics, optoelectronics, sensors, flexible electronics, memory, spin and valleytronics \cite{xu2014spin,xiao2012coupled,shrivastava2021roadmap,fiori2014electronics,rai2018progress,mak2016photonics,chaudhary2016potential,schwierz2011flat,nourbakhsh2016mos2,li2014photodiode,withers2015light,perea2014cvd,wu2015monolayer,he2012fabrication,ge2018atomristor}. Diverse layer-dependent electronic and optical properties of these material systems make them potential candidates for beyond-CMOS applications in emerging 2D nanoelectronics and optoelectronics \cite{shrivastava2021roadmap,liu2022uniform}.  
These materials can be thinned down to monolayer limit, enabling higher surface-to-volume ratios, tuning the bandgap and accordingly the electronic and optical properties, thereby achieving a better electrostatic gate control and carrier confinement over its bulk counterparts \cite{Das2015}.
The large family of TMDs have a wide range of bandgaps, thus when different 2D materials are combined it can lead to many more exciting properties, in addition to those of the individual 2D materials \cite{Novoselov2016}.

Demonstrations of scalable device applications and realization of the potential of these TMD materials to the fullest, however, have been hindered primarily due to the lack of repeatable, reliable and adaptable thin-film synthesis techniques. 
So far, high-performance devices have been reported on materials generally obtained from the top-down micromechanical exfoliation of flakes from the bulk \cite{das2013high,akinwande2019graphene}.
However, this method is not the realistic approach for scalable implementation. For practical applications, various bottom-up approaches, for example, molecular beam epitaxy (MBE) \cite{yue2017nucleation,jiao2015molecular,roy2016structural} and chemical vapor deposition (CVD) \cite{chowdhury2020two,cai2018chemical}, have been utilized to obtain large-area film. While MBE offers high-quality crystalline films with a precise control over the thickness, there are limiting factors like smaller grain size and defects in the grown film. On the other hand, CVD has produced crystalline film with larger grain sizes with the properties that are comparable to the exfoliated flakes. 
However, large-area film with uniform thickness is still difficult to reproduce in CVD due to the lack of precise control over the growth parameters \cite{dong2019kinetics}.
Atmospheric pressure CVD (APCVD) growth produces TMD films with varying thickness with the thickness reducing away from the center of the substrate. Monolayer regions are formed only at the periphery. For a better control over the uniformity, it is important to understand the spatial variation of the density profile and critical roles of the factors influencing this variation.

Thus, it would be useful to find a simple way to theoretically or computationally investigate and predict the deposition profiles at steady states, under the effect of the variations in external parameters like the speed of precursor gas, diffusivity of the deposited particles, temperature, etc.

In literature, the asymmetric simple-exclusion process (ASEP) model, with TASEP (totally asymmetric simple-exclusion process) as its special case, has been used extensively as a model for protein synthesis \cite{MacDonald1969}, Markov processes \cite{Spitzer1970,Liggett1985,Spohn1991}, transport of large molecules through narrow pores \cite{Levitt1973}, traffic flow (single or multi-laned) \cite{Schreckenberg1996,Sopasakis2006,Hilhorst2012}, etc. For a detailed review of the literature of exclusion processes and the methods of solution, see 
\cite{Mallick2006,Derrida1997,Derrida1998,krapivsky2010kinetic} and the references therein. The scaling behaviours in growth processes were studied via the one-dimensional ASEP model in \cite{Krug1997,Zhang1995}. The application of Bethe Ansatz to obtain solutions for such models was outlined in \cite{dhar1987exactly}.
In this article, the 2D ASEP is employed to model the coarse-grained density profile of precursor materials in a CVD growth process, thereby determining the overall uniformity of the deposited film.

Typically, the methods used to study such systems are the density functional theory (DFT) \cite{Argaman2000,Ageev2016} or the kinetic Monte Carlo (KMC) methods \cite{Levi1997,Andersen2019}). The ASEP model, where the lattice is being studied at a coarse-grained level, is computationally faster and less resource-intensive than DFT. It also allows for an easier visualization of the growth process as compared to KMC.

Since in the system that has been considered, the carrier gas acts as a driving force during the deposition process, it is unlikely that a detailed balance relation among the hopping rates would be satisfied. The following processes take place simultaneously: (i) flow of precursor particles along with the carrier gas, (ii) deposition of particles on the substrate, and (iii) diffusion of particles on the surface of the substrate. 
It was found that the ASEP model is a good indicator of the coarse-grained steady-state as well as transient density profiles.  

Our method can be used to model monolayer growth (as in this article), and should be generalizable to multilayers.  The model would be able to predict the effects of relative changes in parameters on the growth in an experimental setup. Compared to Kinetic Monte Carlo simulations, this method is better suited to the simulation of a flow lattice, which has been used to model the depositions on the substrate lattice. Our method can also simulate a much larger sample size, at a coarse-grained level, for larger time scales as compared to simulations based on the Density Functional Theory and Molecular Dynamics.


In Sec. \ref{sec:Expt}, a basic experimental setup of the CVD experiment is discussed. Sec. \ref{sec:Tasep2D} formulates the ASEP model in two dimensions (2D) with (partially) open boundaries, and introduces the concept of `flow' and `growth' in simulation. 
Sec. \ref{sec:Results}  examines various growth profiles and their dependence on several parameters in 2D, ending with a brief discussion on the extension of the model to three dimensions (3D).
Lastly, Sec.\ref{sec:Conclusion} concludes our discussions, and in appendix \ref{sec:1D_TASEP}, the agreement between numerical simulations of the 1D TASEP model and analytical results is shown.

\section{\label{sec:Expt}Experimental setup}

\begin{figure*}
\includegraphics[width=0.9\linewidth]{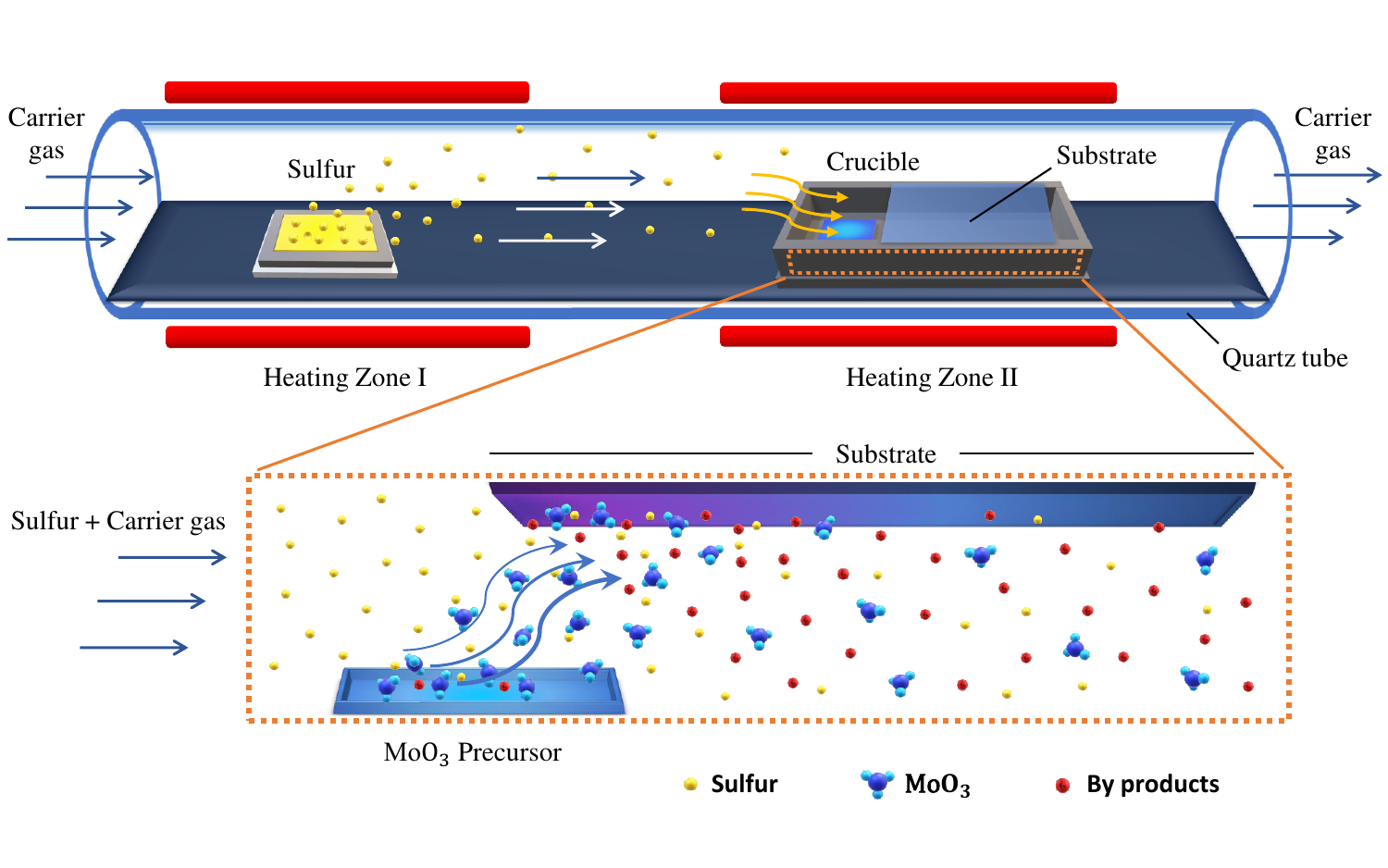}
\caption{\label{fig:SchematicDiagram}{\textbf{Schematic of the experimental setup used for the growth of monolayer TMDs on a substrate:} For MoS$_2$ growth, sulphur and MoO$_3$ precursors are placed in a crucible in the furnace tube and the Si/SiO$_2$ substrate is placed just to the right of the MoO$_3$ precursor. The carrier gas is flown from left to right inside the chamber at a high velocity, enabling the precursors to chemically combine, reach the substrate and deposit onto it.}}
\end{figure*}

A schematic diagram of the experimental setup, where the growth of MoS$_2$ on Si/SiO$_2$ substrate has been carried out via CVD, is shown in Fig. \ref{fig:SchematicDiagram}. Growth of 2D MX$_2$ (M=Mo, W, etc.; X=S, Se, etc.) is carried out in a single zone furnace for the CVD growth under atmospheric pressure. Typically, the growth is conducted on a substrate (e.g., Si/SiO$_2$) using solid precursors (e.g., MoO$_3$, WO$_3$) and chalcogen (e.g., S, Se) powder with Ar/N$_2$/H$_2$ (or a mixture of these gases) as the carrier gas. The substrate is placed on an alumina combustion boat/crucible that contains the metal precursor, which is then positioned at the center of the quartz tube \cite{chowdhury2020two, cai2018chemical}. Another boat containing excess chalcogen powder is kept upstream in the tube, outside the central heating zone, and heated using a separate coil heater. The growth is conducted under chalcogen-rich environment at high temperatures (typical growth temperatures are 850 $^\circ$C and 900 $^\circ$C for MoS$_2$ and MoSe$_2$, respectively). Different characterization techniques, for example, optical microscopy, scanning electron microscopy (SEM), Raman and Photoluminescence (PL) spectroscopies, X-ray photoelectron spectroscopy (XPS), etc. are used to characterize the grown film to confirm the monolayer/multilayer growth, as presented in ref. \cite{chowdhury2020two}.

\begin{figure*}
\includegraphics[width=0.9\linewidth]{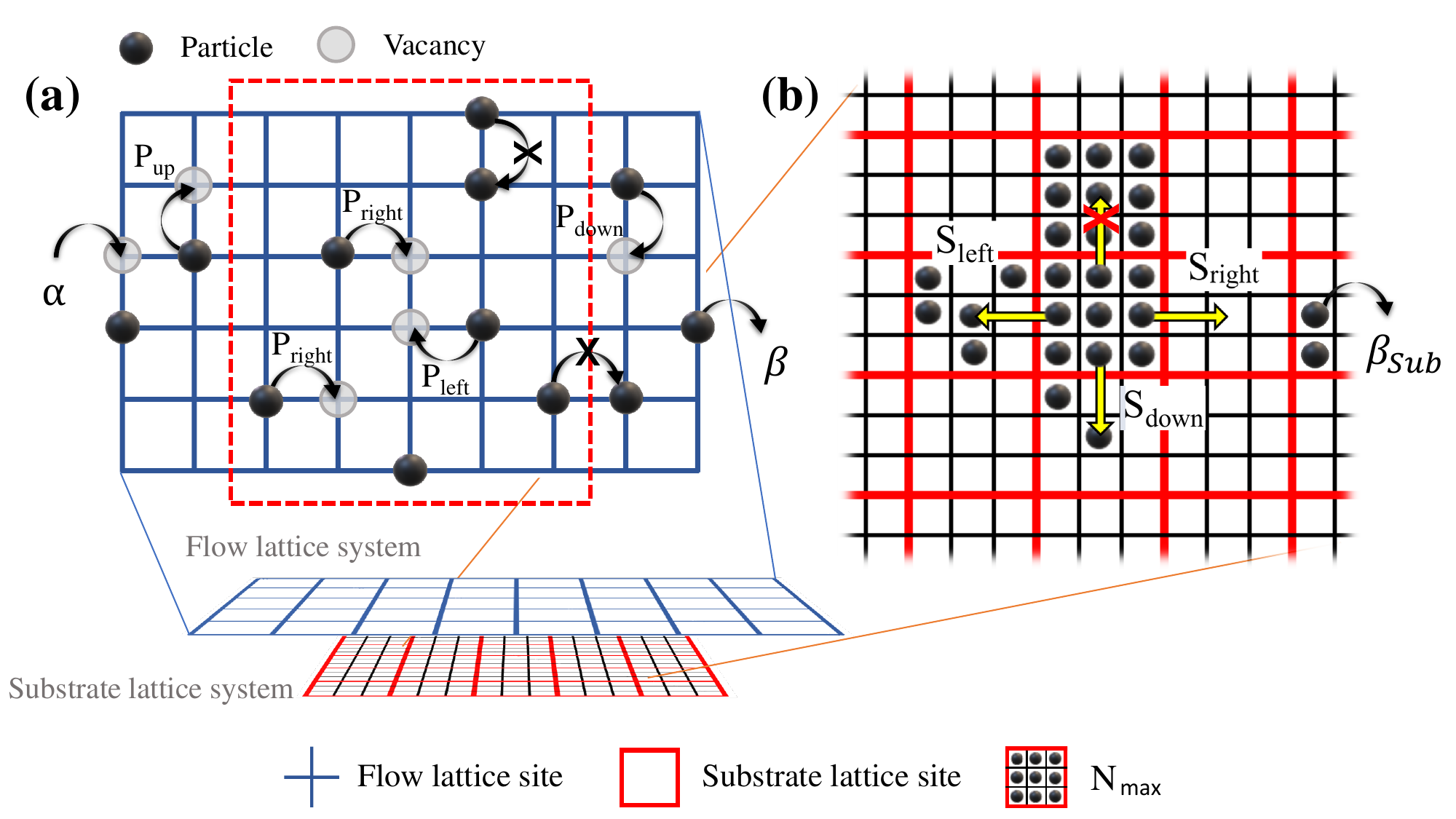}
\caption{\label{fig:Flow+Substrate}{\textbf{Flow lattice and Substrate lattice, showing movement of particles in both:} (a) Grid in blue represents flow lattice. Grid in red represents substrate lattice, with dimensions smaller than that of the flow lattice as shown. The relative positions of the flow system and the substrate surface similar to the experimental setup are sketched. (b) Growth on \textit{each} substrate site can have a maximum value of $N_{\rm max}$. This is shown as a red box full of 9 solid black circles in the sketched example.}}
\end{figure*}

\section{\label{sec:Tasep2D}Model}

The two-dimensional ASEP framework is used to model the flow system as well as the deposition. The accuracy of our simulations is established by demonstrating the agreement with analytical results of the 1D TASEP (discussed briefly in appendix \ref{sec:1D_TASEP}). Below, using Monte Carlo techniques similar to the 1D model, its generalization to two dimensions is discussed.
For computational efficiency, we have considered a multiscale model which consists of two 2D lattice systems, the flow lattice and the growth lattice. The flow lattice system has a relatively larger scale than the growth lattice system. This has been shown in Fig. \ref{fig:Flow+Substrate} (explained in further details below).

To keep the results scale-independent, the variables are in dimensionless units. 
The positions are normalized by the grid size, while the times are normalized by the hopping time or one Monte Carlo step.

\textbf{Flow system:}
In view of the experimental design, the carrier gas flows from left to right in our model (see Fig. \ref{fig:SchematicDiagram}).
The flow system is considered to be a 2D $M\times M$ square lattice onto which particles enter at a rate $\alpha$ through the allowed sites on the left edge (if empty). On the other hand, the particles exit the lattice at a rate $\beta$ through the allowed sites on the right edge (if occupied). 

Initially, all sites on this lattice are unoccupied, i.e., they are initialized as 0.
We assume the gas to have a low concentration, where this effect is significant. At higher concentrations, this contribution will be negligible as compared to deposition of particles.
The size of the injection window is related to the separation between the metal precursor and the substrate (see Fig. \ref{fig:SchematicDiagram}). A smaller injection window corresponds to a small number of particles entering the flow lattice (considered to be the layer closest to the substrate surface) within a time step, implying that the above separation is smaller. Conversely, a larger injection window indicates a larger separation. 
Within the lattice boundaries, for simplicity, hard-core exclusion is imposed, implying that not more than one particle can occupy a given site on the lattice. A schematic presentation of the model has been shown in Fig. \ref{fig:Flow+Substrate}(a). Each particle (except at the edges) on the flow lattice can move in all four directions with probabilities given by $P_{\rm left},~P_{\rm right},~P_{\rm up}$, and $P_{\rm down}$, respectively, with the subscripts denoting the hopping directions.
The horizontal flow of carrier gas necessitates $P_\mathrm{left}\ne P_\mathrm{right}$, while $P_\mathrm{up}=P_\mathrm{down} = 1/4$. The \emph{right bias} $b_\mathrm{right}$ acting on the particle is defined as the difference in the hopping probabilities $P_\mathrm{right}$ and $P_\mathrm{left}$:
\begin{eqnarray}
b_\mathrm{right} = P_\mathrm{right} - P_\mathrm{left}
 \label{eq:RightBias}
\end{eqnarray}

\textbf{Deposition:}
 In the presence of sufficient kinetic energy and substrate temperature, deposited particles will undergo diffusion on the substrate as well, although to a lesser extent than the gaseous flow system. For simplicity, the desorption of the deposited material from a bulk site on the surface is ignored.
Such a condition is satisfied if the substrate temperature is low enough \cite{Barabasi1995}.

To implement this model, we define a substrate area $X\times Y$, where $X, Y \leq M$, with $M$ being the flow lattice dimension, $X$ being the horizontal substrate dimension, and $Y$, the vertical substrate dimension  (see Fig. \ref{fig:Flow+Substrate}(a)). Initially, the entire substrate is completely unoccupied.
Deposition of a precursor particle can occur only when it is within substrate area (marked by the red dashed rectangle in Fig. \ref{fig:Flow+Substrate}(a)), else it remains in the flow system. 
The deposition rates at a given coarse-grained site is assumed to depend on the occupancy of that site. The maximum deposition rate is defined as the deposition rate at a completely unoccupied site on the substrate.
When a particle is over the site $(i,j)$ of the substrate lattice, it can get deposited at this site  with a rate

\begin{eqnarray}
R_{ij} = R_{\rm max} \left(1-\frac{N_{ij}}{N_{\rm max}}\right).
 \label{eq:Rij}
\end{eqnarray}

Here, $N_{ij}$ is the growth at the site $(i,j)$. 
$N_{\rm max}$ refers to the number of actual sites corresponding to a single site on the coarse-grained lattice (see the red squares in Fig. \ref{fig:Flow+Substrate}(b)). Thus, even though such a coarse-grained site can accommodate multiple particles, at finer levels an individual site can accommodate only a single particle, corresponding to a monolayer growth. 
 In the example shown in the Fig. \ref{fig:Flow+Substrate}(b), the highest growth on a site is $N_{\rm max}=9$ (in our simulations, $N_{\rm max}=100$). 
$R_{\rm max}$ is maximum rate of deposition of the precursor particles to a site. Note that the above form of $R_{ij}$ is qualitatively in agreement with the expected decrease in deposition rate with the increase in growth, at a given site.

\textbf{Diffusion:}
A deposited particle can undergo diffusion on the substrate surface.
This process is controlled by a parameter $D$ such that $0 \leq D < 1$. Thus, now the associated hopping probabilities get reduced by a factor of $D$.
Subsequently, the probability to hop to a neighboring site becomes:
\begin{subequations}
  \begin{align}
      S^{\rm right}_{ij} &= D   P_{\rm right}   \left(\frac{N_{ij}}{N_{\rm max}}\right)   \left( 1 - \frac{N_{i+1,j}}{N_{\rm max}}\right); \\
      S^{\rm left}_{ij} &= D   P_{\rm left}   \left( \frac{N_{ij}}{N_{\rm max}}\right)   \left(1 - \frac{N_{i-1,j}}{N_{\rm max}}\right); \\
      S^{up}_{ij} &= D   P_{\rm up}   \left( \frac{N_{ij}}{N_{\rm max}} \right)   
      \left( 1 - \frac{N_{i,j+1}}{N_{\rm max}} \right); \\
      S^{\rm down}_{ij} &= D P_{\rm down}   \left( \frac{N_{ij}}{N_{\rm max}} \right)   \left( 1 - \frac{N_{i,j-1}}{N_{\rm max}}\right).
  \end{align}
        \label{eq:HoppingProbabilitiesSubstrate}
  \end{subequations}
The subscripts and superscripts of $S$ denote the site under consideration and the hopping direction, respectively. 
The set of equations can be compactly written as:
\begin{align}
    S(i',j'|i,j) = DP(i',j'|i,j) p(i,j) q(i',j'),
\end{align}
where $S(i',j'|i,j)$ is the probability of hopping from site $(i,j)$ to site $(i',j')$ on the lattice; $P(i',j'|i,j)$ are the probabilities $P_{\rm left}$, etc.; $p(i,j)$ is the occupation probability $N_{ij}/N_{\rm max}$ of the site $(i,j)$, and $q(i',j')$ is the probability $(1-N_{i'j'}/N_{\rm max})$ of a vacancy at the site $(i',j')$.

The above forms are consistent with the expected \textit{qualitative} behaviour of the hopping probability. Higher the growth on a given site, higher will be the tendency of a deposited particle to hop away from that site, while a higher growth on an adjacent site will reduce the tendency to hop into it. 
This follows from the fact that a relatively filled sublattice (see Fig. \ref{fig:Flow+Substrate}) corresponding to that coarse-grained site will obstruct the hopping of particles from a neighboring sublattice due to the simple-exclusion interactions. For the same reason, a particle in a filled sublattice will try to hop out of it.
Note that there is a finite probability of the deposited particles not diffusing from the site of deposition, given by $1-(S_{ij}^{\rm left}+S_{ij}^{\rm right}+S_{ij}^{\rm up}+S_{ij}^{\rm down})$.

It readily follows from Eq. \eqref{eq:HoppingProbabilitiesSubstrate} that the probabilities of diffusion on the substrate from a site that is full (i.e., $N_{ij}=N_{\rm max}$) to the adjacent site on the right, if vacant, is given by $S^\mathrm{right}_{ij} = D~P_\mathrm{right}$. Similar equations can be written for the other three directions. The ejection probability for each site on the rightmost edge of the substrate is kept as $(N_{Xj}/N_{\rm max})\beta_{\rm sub}$, where $\beta_{\rm sub}$ is the rate of ejection from the substrate when such a site is fully occupied ($N_{Xj} = N_{max}$), and the equation $i = X$ describes the right edge.

A summary of the various parameters investigated and the factors that affect them has been provided in appendix \ref{sec:parameters}.

\section{Algorithm}

A unit Monte Carlo step in the simulations implements the following:
\begin{enumerate}
    \item Injection into flow system: Each empty site of the injection window at the left edge is filled individually with a probability $\alpha$.
    \item Exclusion: Each particle within the flow lattice can hop to an adjacent site with the relevant probability (among $P_{\rm left},~P_{\rm right},~P_{\rm up}$ and $P_{\rm down}$), provided the latter is empty.
    \item Determination of deposition rates: $R_{ij}$ is determined (from Eq. \eqref{eq:Rij}) for each site of the flow lattice, that lies vertically above a substrate lattice site.
    \item Each particle within the substrate area in the flow lattice is deposited onto substrate site at the same position with probability $R_{ij}$.
    \item The hopping probabilities $S^{\rm left}_{ij}$, $S^{\rm right}_{ij}$, $S^{\rm up}_{ij}$ and $S^{\rm down}_{ij}$ are determined from Eq. \ref{eq:HoppingProbabilitiesSubstrate}.
    \item Diffusion on substrate: A particle from each substrate site (red squares shown in Fig. \ref{fig:Flow+Substrate}) lattice can hop to adjacent site with the directional probabilities appearing in step 5.
    \item Ejection from substrate: A particle from the right edge of the \textit{substrate} lattice is ejected with probability $(N_{Xj}/N_{\rm max})\beta_{\rm sub}$.
    \item Ejection from flow system: Each particle on rightmost edge of the \textit{flow} lattice is ejected with probability $\beta$.
\end{enumerate}

\section{\label{sec:Results}Results and discussions}

\subsection{\label{subsec:Deposition} Growth profiles in 2D}

Here we study the growth profile on a substrate using the 2D ASEP model by imposing a finite deposition rate of the precursor particles on the substrate, and supplementing it by including the effect of diffusion following the deposition.  
The kinetic energy of the deposited particles would be manifested through their diffusion rate on the substrate.

A substrate lattice with dimensions $60 \times 100$ is considered, and placed over the flow lattice of dimensions $100 \times 100$ (see Fig. \ref{fig:Flow+Substrate}). The substrate is placed across sites $i=21$ to $i=80$ along the $x$-axis (corresponding to $i=1$ to $i=60$ in the coordinates of the substrate lattice), and $j=1$ to $j=100$ along the $y$-axis. Variation in the growth profile is studied as a function of different growth parameters.
The entry sites are restricted symmetrically from $j=46$ to $j=55$ on the left edge \textit{of the flow lattice}.  The maximum allowed growth at any site on the substrate lattice is set to $N_{\rm max}=100$. 
As mentioned in Sec. \ref{sec:Tasep2D}, the time $t$ is in units of Monte Carlo time steps throughout this article.

\subsubsection{Dependence on the time of evolution}

\begin{figure*}[!h]
\includegraphics[width=1\linewidth]{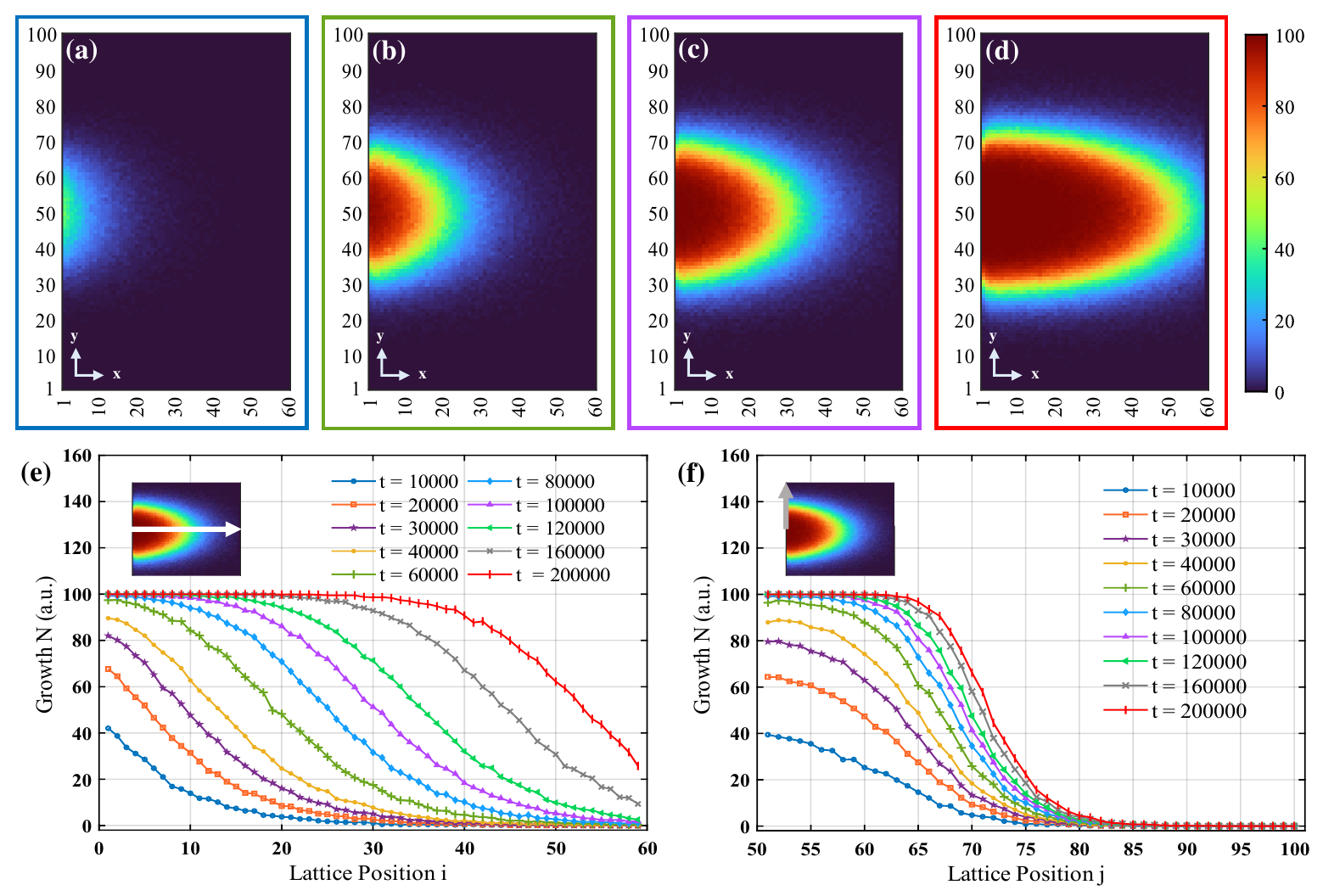}
\caption{\label{fig:SubstrateWithTime}{\textbf{Dependence of 2D growth profiles on time of evolution:} The growth profiles are taken at times (a) $t = 10^4$, (b) $t = 6\times 10^4$, (c) $t = 10^5$, and (d) $t = 2\times 10^5$. All profiles and plots are obtained for a flow lattice of dimensions $100 \times 100$, with  right bias $b_\mathrm{right} = 0.2$ and $\alpha = \beta = 1$ and an injection window ranging from sites 46 to 55 on the left edge of flow lattice, and an ejection window ranging from sites 1 to 100 on its right edge. 
The substrate covers the flow lattice sites 1 to 100 along $y$-axis and sites 21 to 80 along $x$-axis. Other parameters are $N_{\rm max} = 100, ~R_{\rm max} = 0.03, ~D = 0.005$, and $\beta_{\rm sub}=1$. Each profile is obtained by ensemble
averaging over 10 realizations. 
The colors of the outer borders imply that the snapshots have been taken at the values of time corresponding to the plots with the same colors appearing in the lower panel.
(e) is a plot between growth $N$ and lattice position (sites 1 to 60) in the $x$-direction for different values of $t$. Similar profiles along the $y$-direction (for sites 50 to 100) are shown in (f).
}}
\end{figure*}

\begin{figure*}[htbp!]
\includegraphics[width=1\linewidth]{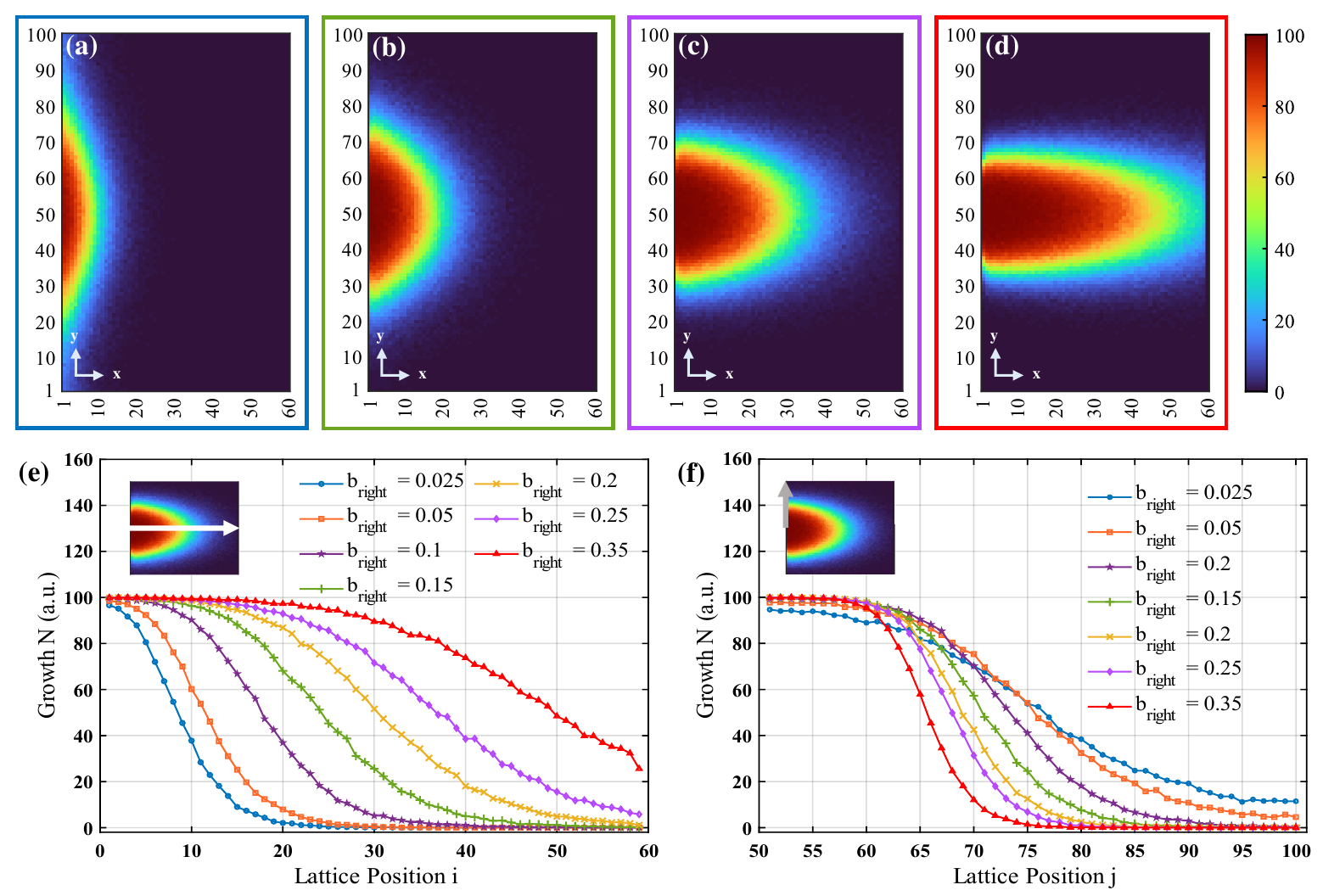}
\caption{\label{fig:SubstrateWithRB}{\textbf{Dependence of 2D growth profiles on $b_{\rm right}$:} The growth profiles are shown for (a) $b_\mathrm{right}=0.025$, (b) $b_\mathrm{right}=0.1$, (c) $b_\mathrm{right}=0.2$, and (d) $b_\mathrm{right}=0.35$. All profiles and plots are obtained at time $t = 10^5$.  All other parameters are kept the same as in Fig. \ref{fig:SubstrateWithTime}. 
Each profile is obtained by ensemble averaging over 10 realizations. 
The colors of the outer borders imply that the snapshots have been taken at the values of $b_{\rm right}$ corresponding to the plots with the same colors appearing in the lower panel.
(e) is a plot between growth $N$ and lattice position in the $x$-direction for different values of $b_{\rm right}$. Similar plots along the $y$-direction are shown in (f).
}}
\end{figure*}

 Fig. \ref{fig:SubstrateWithTime}(a) to (d) show the density profiles of the depositions at different evolution times (see figure caption). 
Figs. \ref{fig:SubstrateWithTime}(e) and (f) show the variation of growth along the $x$ and $y$ axes, respectively (along the arrows shown in the insets) for different values of $t$. In Fig. \ref{fig:SubstrateWithTime}(e), it is observed that with increase in the time of evolution, the plateau region near the left edge increases, while the same near the right edge decreases. Qualitatively similar trend was observed  in the experimental results as well - the deposition becomes more uniform at longer times \cite{chiawchan2021cvd,wang2014shape,you2018synthesis}.
Analogous changes are observed along the $y$-direction (Fig. \ref{fig:SubstrateWithTime}(f)), with the exception that in all cases there is a large drop in the density as we move away from the source. This is due to the rightward current caused both by the finite right bias and the values of injection and ejection rates. For the set of parameters chosen, the growth becomes negligible beyond site $j=85$ along the $y$-axis.

\subsubsection{Dependence on Right Bias}

Fig. \ref{fig:SubstrateWithRB} compares the deposition after a fixed time of observation ($t=10^5$) for different values of right bias. From (a) to (d), with increase in $b_\mathrm{right}$, the growth profile on the substrate gets stretched out rightward. The qualitative nature of Fig. \ref{fig:SubstrateWithRB}(e), where the growth profiles along the $x$-axis have been plotted, closely resembles that of Fig. \ref{fig:SubstrateWithTime}(e). Fig. \ref{fig:SubstrateWithRB}(f), however, exhibits substantial qualitative difference from Fig. \ref{fig:SubstrateWithTime}(f). In the latter case, the densities close to the source were different for different times of evolution, although they merged into a single line a bit further. In this case, the values of growth densities close to the source are almost the same in all cases, owing to the fact that the parameters $R_{max}$ and $D$ are the same for all profiles. 
With increase in $b_{\rm right}$, the curve decays faster, indicating a smaller divergence of the deposition along the $y$-axis. The reason is that, higher the right bias, smaller is the probability of the particle wandering off in the direction perpendicular to the direction of the bias.
The qualitative agreement of our results with those of \cite{chiawchan2021cvd,tummala2020application} can be readily observed.

\subsubsection{Dependence on $R_{\rm max}$}

\begin{figure*}
\includegraphics[width=1\linewidth]{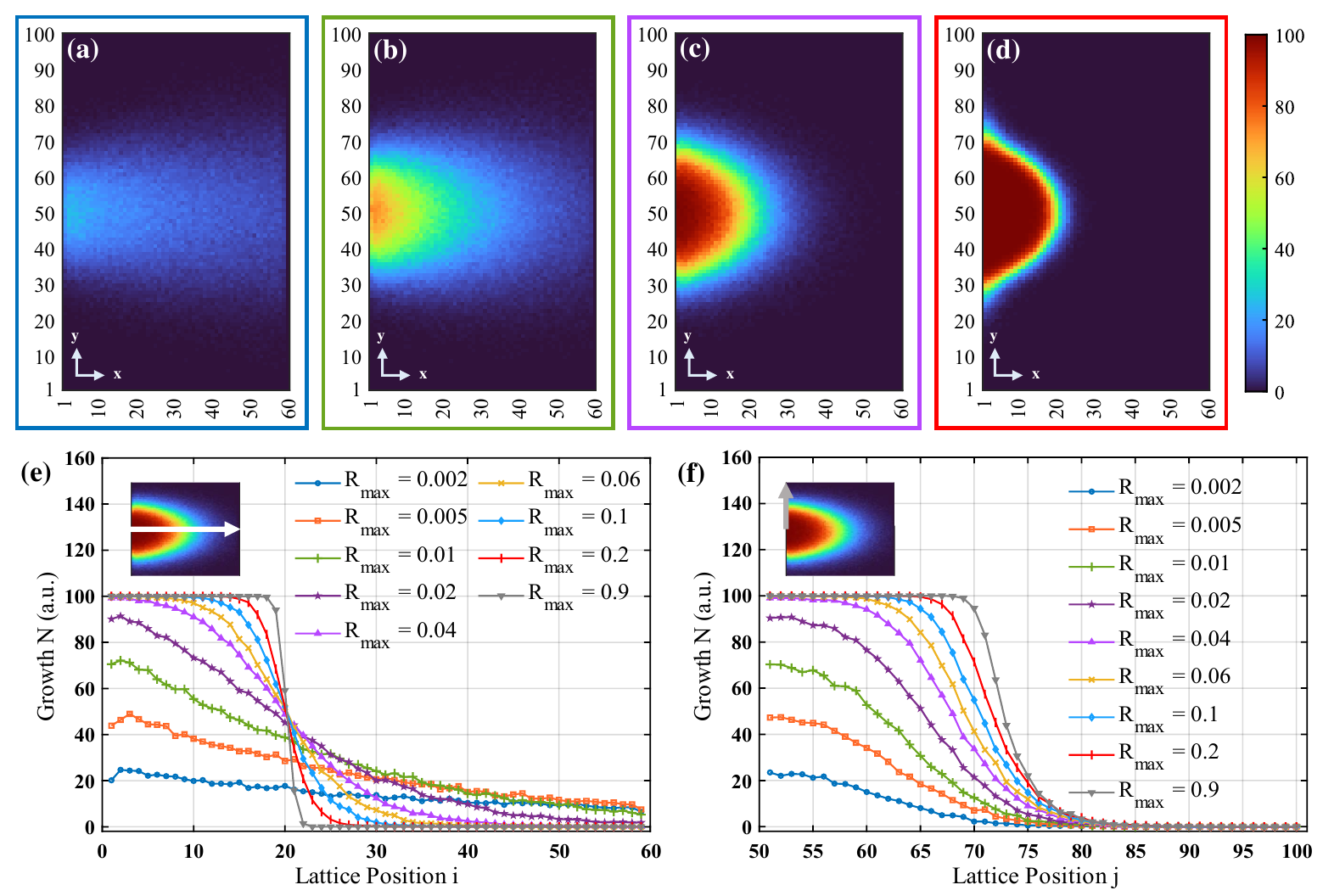}
\caption{\label{fig:SubstrateWithMDR}{\textbf{2D growth profiles for different values of $R_{\rm max}$:} The growth profiles are shown for (a) $R_{\rm max}=0.002$, (b) $R_{\rm max}=0.01$, (c) $R_{\rm max}=0.04$, and (d) $R_{\rm max}=0.2$. We set $b_\mathrm{right} = 0.2$ and $t=6\times 10^4$, while all other parameters are kept the same as in Fig. \ref{fig:SubstrateWithTime}. Each substrate plot is an average of 10 realizations. 
The colors of the outer borders imply that the snapshots have been taken at the values of $R_{\rm max}$ corresponding to the plots with the same colors appearing in the lower panel.
(e) is a plot between $N$ and the lattice position in $x$ direction at different times. Similar plots along the $y$-direction are shown in (f).}}
\end{figure*}

Fig. \ref{fig:SubstrateWithMDR} shows growth profiles for several values of $R_{\rm max}$ (see Eq. \eqref{eq:Rij}). The observation is recorded at time $t = 6\times 10^4$, with the other parameters as mentioned in the figure caption.   Thinning out of the depletion region (where the growth is intermediate between the maximum and the minimum values) with increasing $R_{\rm max}$ (Fig.   \ref{fig:SubstrateWithMDR}(a) to (d)) is readily observed. 
This is also clear from Fig. \ref{fig:SubstrateWithMDR}(e) and (f), showing depositions along the $x$ and $y$ directions respectively, where the curves decay faster for higher $R_{\rm max}$. Accordingly, the grown domains will attain sharper boundaries \cite{chowdhury2020two}.
The occupancy remains constant for a larger $x$-range for higher values of $R_{\rm max}$.
The maximum occupancy itself (intercept with the growth axis) is found to increase with $R_{\rm max}$, and reaches $N_{\rm max}$ for $R_{\rm max}\gtrsim 0.04$. For very low $R_{\rm max}$, the combined effect of $\alpha$, $\beta$ and $D$ will give rise to an almost uniform deposition. If $R_{\rm max}$ is then increased, the sites close to the left edge will fill up faster due to $\alpha$, while the sites close to right edge will deplete faster due to $\beta$.
On the other hand, in the $y$-direction (subfigure (f)), although the decay is faster for higher $R_{\rm max}$, the curves merge together beyond site $j=80$, instead of intersecting.

\subsubsection{Dependence on Diffusion}

\begin{figure*}
\includegraphics[width=1\linewidth]{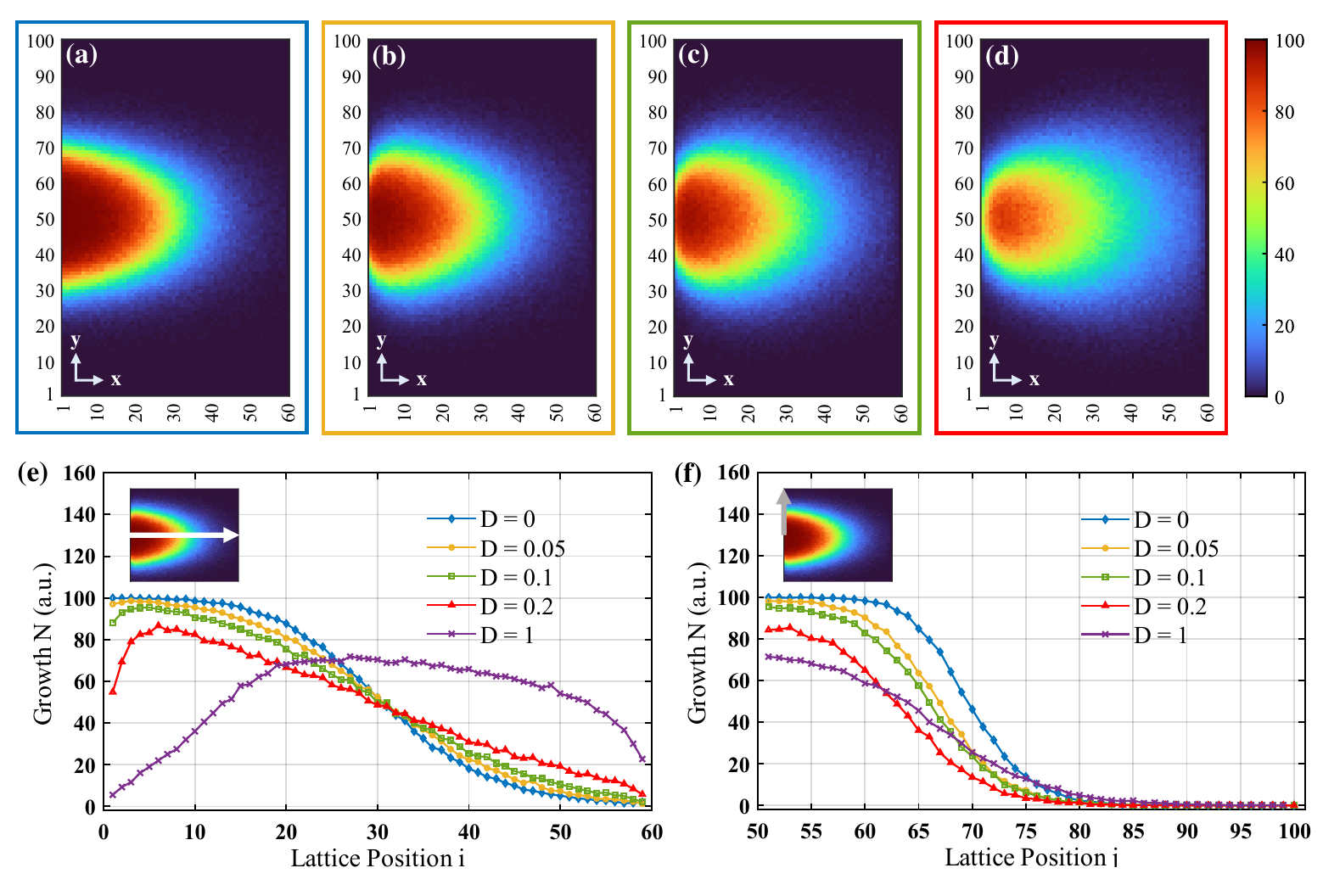}
\caption{\label{fig:SubstrateWithD}{\textbf{Effect of diffusion on the 2D growth profiles:} 
The growth profiles are shown for (a) $D=0$, (b) $D=0.05$, (c) $D=0.1$, and (d) $D=0.2$. We set $t = 1\times 10^5$, while all other parameters are kept the same as in Fig. \ref{fig:SubstrateWithTime}. Each substrate plot is an average of 10 realizations. 
The colors of the outer borders imply that the snapshots have been taken at the values of $D$ corresponding to the plots with the same colors appearing in the lower panel.
(e) is a plot between $N$ and the lattice position in $x$ direction with varying $D$. Similar plots along the $y$-direction are shown in (f).}}
\end{figure*}

Fig. \ref{fig:SubstrateWithD}(a) to (d) show the growth profiles for increasing values of $D$.
An increase in the depletion region width is correspondingly observed, along with a reduction in the maximum value of the occupancy $N$. This is because, with increasing diffusion on substrate, there is a greater tendency of particles to move from higher to lower density. This is also evident from fig. \ref{fig:SubstrateWithD}(e). With increasing $D$, the maximum $N$ decreases, and the maxima shift rightward. At the same time, the plateau-like shape (close to the left edge) changes to a non-monotonic one with a distinct maximum.  
In Fig. \ref{fig:SubstrateWithD}(f) as well, the maximum $N$ decreases with increasing $D$. However, the curves remain monotonic, which is expected from the symmetry of the setup.

\begin{figure*}
\includegraphics[width=1\linewidth]{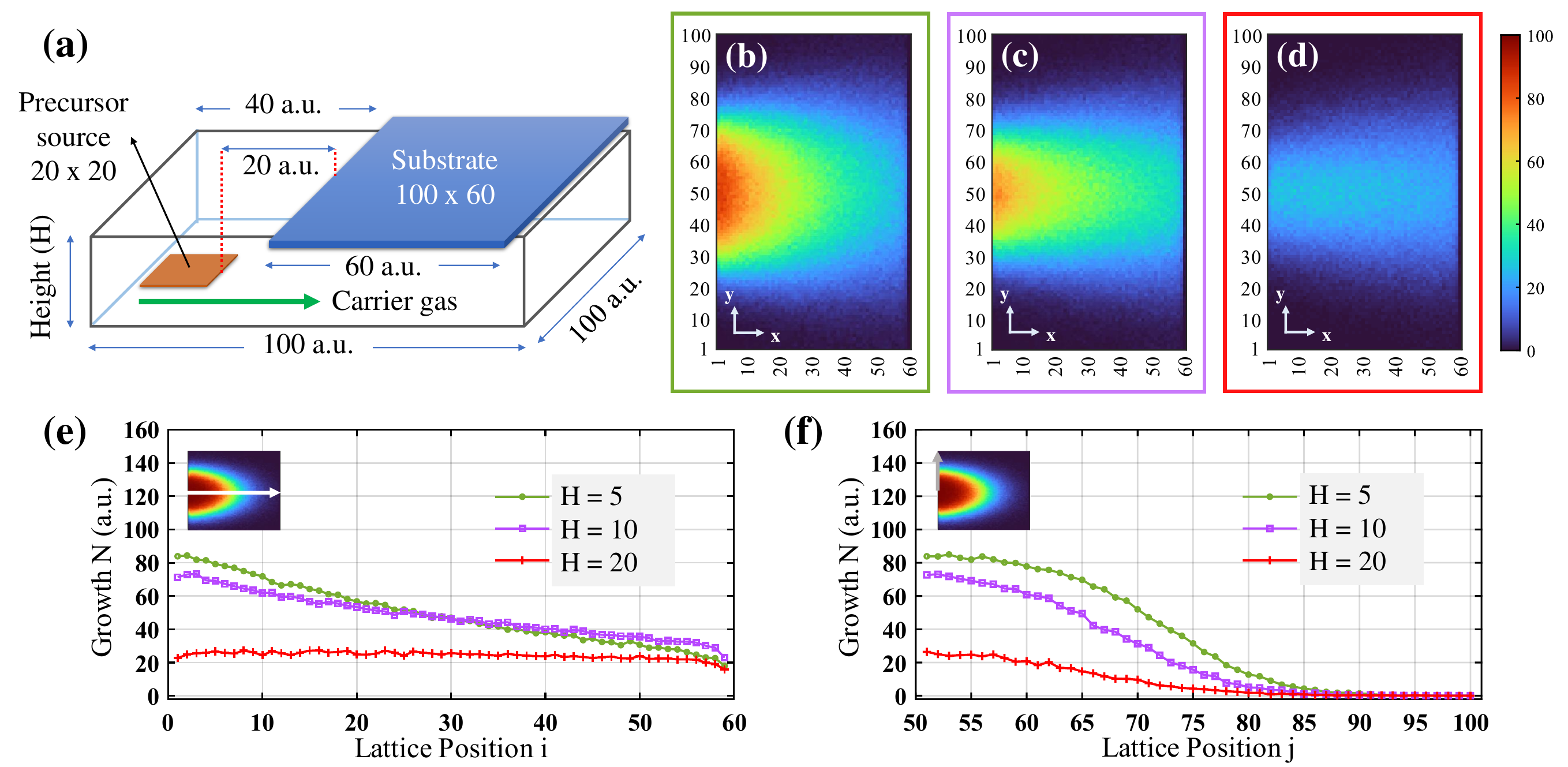}
\caption{\label{fig:SubstrateWithHeight}{\textbf{Growth profiles for 3D ASEP for different heights between precursor and substrate} (a) displays the schematic of 3D ASEP with position of substrate. All profiles and plots are obtained for a flow lattice of dimensions $100 \times 100 \times H$, and growth profiles are taken at heights (b) $H = 5$, (c) $H = 10$, and (d) $H = 20$, with horizontal space between the precursor and substrate being 20 arbitrary units. Right bias $b_\mathrm{right} = 0.266$, $\alpha =0.03$, $\beta =0.2$, injection window ranges from sites 46 to 55 on the left edge of flow lattice, and ejection window ranges from sites 1 to 100 on its right edge. 
The substrate covers the flow lattice sites 1 to 100 along $y$-axis and sites 21 to 80 along $x$-axis. Other parameters are $N_{\rm max} = 100, ~R_{\rm max} = 0.08, ~D = 0.03$, and $\beta_{\rm sub}=1$. All profiles and plots are obtained at time $t = 10^4$. Each profile is obtained by ensemble
averaging over 10 realizations. 
The colors of the outer borders imply that the snapshots have been taken at the values of height corresponding to the plots with the same colors appearing in the lower panel.
(e) is a plot between growth $N$ and lattice position (sites 1 to 60) in the $x$-direction for different values of $t$. Similar profiles along the $y$-direction (for sites 50 to 100) are shown in (f).
}}
\end{figure*}

\subsection{\label{sec:3Dtasep} Growth profiles in 3D }

  In addition to the four probabilities of particle hopping in 2D ($P_{\rm left},~P_{\rm right},~P_{\rm up}$, $P_{\rm down}$), a particle in 3D flow lattice has two more probabilities of moving in z-direction, given by  $P_{\rm top}$, and $P_{\rm bottom}$, respectively, with the subscripts denoting the directions of the hopping.

As depicted in Fig. \ref{fig:SubstrateWithHeight} (a), a 3D flow lattice of size $L\times B \times H$ (here $L=B=100$)  has been considered, onto which particles enter at a rate $\alpha$ through the allowed sites of size $20\times 20$ (if empty) lying vertically above the metal precursor. On the other hand, the particles exit the flow lattice at a rate $\beta$ through the right surface (if occupied). Experimentally, the height $H$ represents the physical height between the metal precursor and the substrate. The latter is of size $100\times 60$ and is placed above the flow lattice (from sites $i=41$ to 100 along $x$-direction, and $j=1$ to 100 along $y$ direction). As before, the particle motion is biased towards the right, and initially, all sites are unoccupied. Particles are allowed to hop to an adjacent site only if it is empty, while the deposition algorithm remains the same as in 2D.

Fig. \ref{fig:SubstrateWithHeight} shows growth profiles for several values of height, corresponding to the parameters mentioned in the figure caption. It is observed that as height increases (Fig. \ref{fig:SubstrateWithHeight}(b) to (d)), the growth area spreads, and the growth becomes more uniform along the $x$ and $y$ directions. 
This is also clear from Fig. \ref{fig:SubstrateWithHeight}(e) and (f), showing depositions along the $x$ and $y$ directions respectively, where the curves decay faster for larger heights.

\section{\label{sec:Conclusion}Conclusions}

In this article, a numerical study of the depositions of transition metal dichalcogenides has been carried out using the ASEP model. 
Despite being simplistic, the model captures several qualitative features  of the coarse-grained density profiles observed in experiments \cite{chiawchan2021cvd,wang2014shape,you2018synthesis,tummala2020application}.
An initial benchmarking of the code used for simulation was carried out by generating the known results for the 1D TASEP model (provided in appendix \ref{sec:1D_TASEP}).
In 2D, the variations in the growth profiles with the time of evolution, right bias, deposition rate, and diffusion rate have been examined.
The general observation is the horizontal stretching of the profiles both with evolution times and right bias. 
With the increase in deposition rate, the profile was observed to approach a step-like function, with a high density close to the left edge. The interplay of this effect with diffusion was manifested in the progressively gradual decay of the function as $D$ is increased relative to $R_{\rm max}$. 
Finally, the 3D case has been briefly visited, where the growth profiles have been studied as a function of the height difference between the precursor source and the substrate surface. The deposition is observed to be more uniform with the increase in this vertical separation.

Our study can be extended further to progressively reduce the coarse-graining, in order to obtain density profiles of higher resolutions. This would help to provide a better control of the uniformity of CVD growth over a large area of 2D materials. 
The model should also be applicable to the study of other 2D materials.


\section{Acknowledgement}

A.R. acknowledges support from the Science and Engineering Research Board, Govt. of India, grant \# SRG/2022/000788.

\appendix
\section{\label{sec:1D_TASEP}Discussion of 1D TASEP model}

We begin by defining the Totally Asymmetric Simple-Exclusion Process for particles, on a one-dimensional lattice model comprised of $N$ sites. Particles are injected from the left edge, and can only hop towards the right (totally asymmetric), provided the site is empty, and then eject when they reach the right edge, both at certain specified rates. No site can contain more than one particle, which means that a particle is not allowed to hop to a site that is already occupied by another particle.

Consider a source of particles on the left, to which the leftmost site of the lattice is connected. Similarly, there is a sink of particles to which the rightmost site of the lattice is connected. The rate at which particles can enter from the source to the first site (provided it is empty) is given by $\alpha$, while the rate at which the particles can leave the $N^\mathrm{th}$ site (provided this site is filled) is given by $\beta$. Note that the lattice sites from left to right are labelled by the site indices $i = 1, 2,..., N$, and the positions are denoted by $x_1,x_2,\cdots,x_N$.

\begin{figure}[!h]
\includegraphics[width=\textwidth]{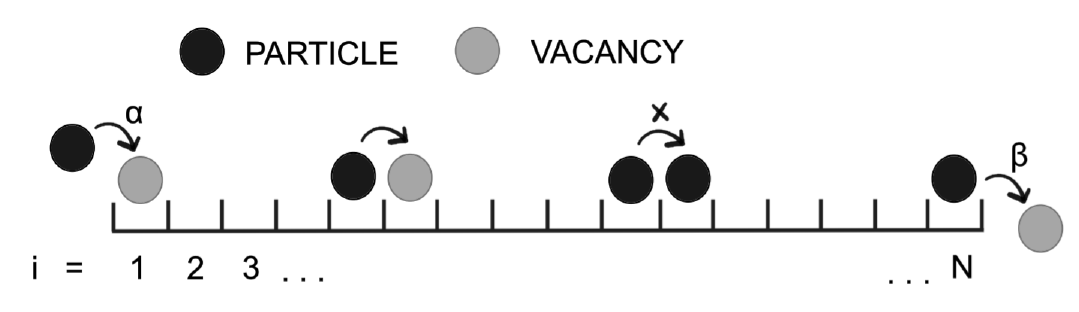}
\caption{\label{fig:1DTasep}{\textbf{TASEP in One Dimension} N site one-dimensional lattice, with site indices marked as i = 1, 2, 3... . Particles are injected into the leftmost site (provided it is vacant), with rate of injection $\alpha$. Similarly, particles are ejected from the rightmost site (provided it is occupied), with rate of ejection $\beta$. In between, in TASEP, particles have probability 1 of hopping rightwards, but they can only hop to the next site provided it is empty, denoted here by a vacancy.}}
\end{figure}

This above discussed system is taken and its time evolution is studied. In a single time step, each of the particles present on the lattice either makes a single hop, if the site to the immediate right is empty, or none at all, if it is occupied. Thus, the configuration of the particles present on the lattice gets updated with every time step. Given that the process is a probabilistic one, the configuration at a given time instant $t$ will in general be different in every different realization of this experiment, even if the initial configuration is chosen to be the same in every case. Thus, a meaningful parameter that can be used to describe the system would be the local density $\rho_i\equiv f(x_i)$, which is a function of the lattice site $x_i$ in general. It is obtained by considering many realizations of the same experiment, and taking an ensemble average of the particle occupancy at each site on the lattice (see \cite{krapivsky2010kinetic}).

The local current of this dynamic system is given by
\begin{eqnarray}
    J_i = \langle \rho_i(1-\rho_{i+1}) \rangle,
\end{eqnarray}
where $\rho_i$ is the occupancy of site $i$ at time $t$ when this current is being computed, and can take values 0 or 1.



\begin{figure*}
  \includegraphics[width=0.9\textwidth]{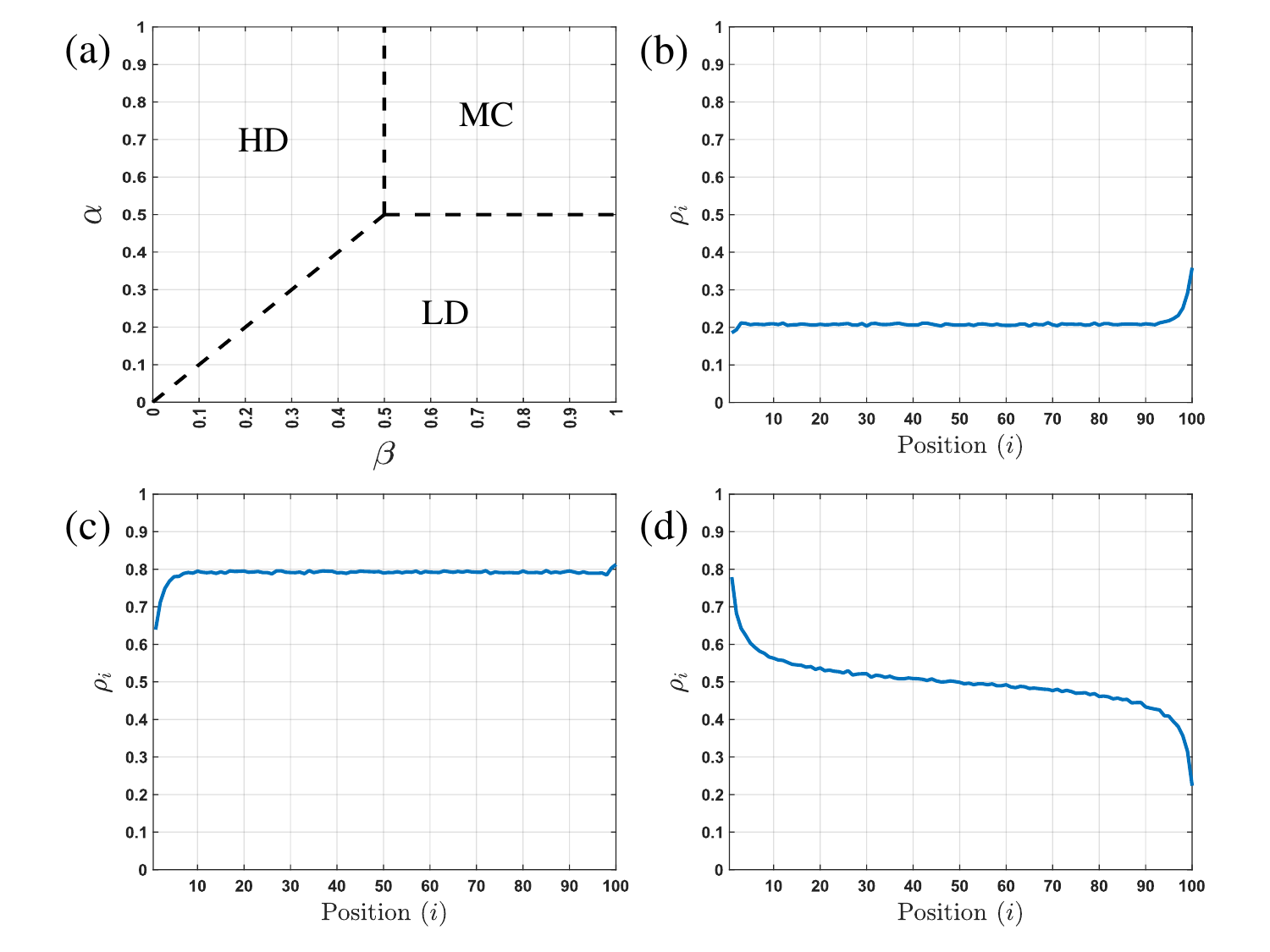}
\caption{\label{fig:OneDimensionSimulations}\small{{\textbf{ Steady State phases for TASEP in One Dimension} Sub-figure (a) depicts and demarcates the various steady state phases that arise in one-dimensional TASEP. Sub-figure (b) shows low density or LD ($\alpha < \beta$) profile, with $\alpha = 0.2$ and $\beta = 0.3$; sub-figure (c) shows high density or HD ($\alpha > \beta$) profile, with $\alpha = 0.4$ and $\beta = 0.2$; and lastly, sub-figure (c) shows maximal current or MC ($\alpha > 0.5,\beta > 0.5$) profile, with $\alpha = 1$ and $\beta = 1$. The last 3 figures have been obtained after 5000 realizations each.}}}
\end{figure*}

The solution to the local density and local current of the TASEP system can be obtained using the matrix-product ansatz \cite{krapivsky2010kinetic}. In particular, the steady state where $\langle J_1\rangle = \langle J_2\rangle = \cdots = \langle J_N\rangle \equiv J$ (say) is considered, where the solutions are given by
\begin{alignat}{3}
    J &= 1/4, \hspace{2cm} && \alpha\ge 1/2, \beta\ge 1/2 \hspace{0.5cm} &&\text{(MC phase)}; \nonumber\\
    &= \alpha(1-\alpha), && \alpha<1/2, \beta >\alpha && \text{(LD phase)}; \nonumber\\
    &= \beta(1-\beta), && \beta<1/2,  \alpha>\beta && \text{(HD phase)}.
\end{alignat}

Correspondingly, the bulk density becomes site-independent: $\rho_m =\rho_{m+1} = \cdots = \rho_n = \rho$ (say) with $m>1$ and $n<N$ (the edge effects are assumed to become manifest for $i<m$ and $i>n$), and are given by
\begin{alignat}{3}
    \rho &= 1/2, \hspace{2cm} && \alpha\ge 1/2, \beta\ge 1/2 \hspace{0.5cm} &&\text{(MC phase)};\nonumber\\
    &= \alpha, && \alpha<1/2, \beta >\alpha && \text{(LD phase)}; \nonumber\\
    &= 1-\beta, && \beta<1/2,  \alpha>\beta && \text{(HD phase)}.
\end{alignat}

Note that there are three phases, indicated by MC (maximal current) phase, LD (low density) phase, and HD (high density) phase.
The agreement between our simulations and the above results is shown in Fig. \ref{fig:OneDimensionSimulations}. The initial condition consists of a completely empty lattice containing 100 sites, and evolution is carried out for 1000 time steps, when the system is found to have reached a steady state. The ensemble averages have been carried out over $5\times 10^4$ realizations of the experiment. Thus, this completes the benchmarking of our codes. 

\section{Summary of the parameters investigated} \label{sec:parameters}
\begin{enumerate}
    \item   $\alpha$ is the rate of entering the flow lattice from the source if the first site is empty. $\beta$ is the rate of leaving the flow lattice if the last site is occupied. These parameters depend on the gas flow that can be externally controlled.
    \item $N_{ij}$ is the occupancy of the sublattice corresponding to a coarse-grained site $(i,j)$. Its maximum value can be $N_{\rm max}$, which is the total number of actual sites corresponding to a given coarse-grained site.
    \item Probabilities of hopping on the flow lattice: $P_{\rm left},~P_{\rm right},~P_{\rm up}$ and $P_{\rm down}$. They are affected typically by the rate of gas influx and temperature. 
    \item The right bias, $b_{\rm right} = P_{\rm right}-P_{\rm left}$, is also affected by the above conditions.
    \item Probabilities of hopping on the substrate lattice: $S^{\rm left}_{ij}$, $S^{\rm right}_{ij}$, $S^{\rm up}_{ij}$ and $S^{\rm down}_{ij}$. They are affected by temperature, since it affects the diffusion rate.
    \item Diffusion parameter $D$, which controls the hopping rates on the substrate. Clearly, this parameter would depend on temperature of the substrate and the density of deposited particles.
    \item Deposition rate $R_{ij}$ at the lattice position $(i,j)$. This is determined by the current occupancy $N_{ij}$ of the site and the maximum possible occupancy $N_{\rm max}$.
    \item The maximum deposition rate $R_{\rm max}$, that decreases with increase in temperature due to enhanced desorption and surface diffusion.
\end{enumerate}

\providecommand{\noopsort}[1]{}\providecommand{\singleletter}[1]{#1}%

\end{document}